\documentclass[manuscript, nonacm, screen]{acmart}
\usepackage[inkscapelatex = false]{svg} 

\AtBeginDocument{%
  }

\setcopyright{acmlicensed}
\copyrightyear{2026}
\acmYear{2026}
\acmDOI{XXXXXXX.XXXXXXX}
\acmISBN{978-1-4503-XXXX-X/2018/06}

\begin{document}

\title{Putting Privacy to the Test: Introducing Red Teaming for Research Data Anonymization}

\author{Luisa Jansen}
\email{luisa.jansen@unibe.ch}
\orcid{0000-0001-8126-1306}
\affiliation{%
  \institution{University of Bern}
  \city{Bern}
  \country{Switzerland}
}

\author{Tim Ulmann}
\email{tim.ulmann@unibe.ch}
\orcid{0009-0005-8082-3254}
\affiliation{%
  \institution{University of Bern}
  \city{Bern}
  \country{Switzerland}
}

\author{Robine Jordi}
\email{robine.jordi@gmail.com}
\orcid{0009-0002-8532-8154}
\affiliation{%
  \institution{University of Bern}
  \city{Bern}
  \country{Switzerland}
}

\author{Malte Elson}
\email{malte.elson@unibe.ch}
\orcid{0000-0001-7806-9583}
\affiliation{%
\institution{University of Bern}
  \city{Bern}
  \country{Switzerland}
}

\renewcommand{\shortauthors}{Jansen et al.}

\begin{abstract}
Recently, the data protection practices of researchers in human-computer interaction and elsewhere have gained attention. Initial results suggest that researchers struggle with anonymization, partly due to a lack of clear, actionable guidance. In this work, we propose simulating re-identification attacks using the approach of red teaming versus blue teaming: a technique commonly employed in security testing, where one team tries to re-identify data, and the other team tries to prevent it. We discuss our experience applying this method to data collected in a mixed-methods study in human-centered privacy. We present usable materials for researchers to apply red teaming when anonymizing and publishing their studies' data. 
\end{abstract}




\maketitle

\section{Introduction and Related Work}
Recent work has shed light on research practices that usually remain hidden: the data protection practices of empirical research data \cite{Martius2025_OutSightOut, Guo2025_HowResearchersDeIdentify}, indicating researchers' insecurities and struggles surrounding data anonymization. With open data gaining more momentum in human-computer interaction (HCI) research, anonymization becomes more important \cite{Jansen2025_TensionOpenData, Oppenlaender2025_MetaHCIFirstWorkshop}: Only with sufficient anonymization can researchers share their data while upholding their participants' privacy and respecting the trust that is at the foundation of the researcher-participant relationship
\cite{Guillemin2018_ResearchParticipantsTrust}. This obligation is structurally embedded in research practice: Ethical codes such as the ACM's \cite{ACMPublicationsBoard2021_ACMPublicationsPolicy} as well as norms and commitments within the HCI community establish anonymization as a standard requirement for responsible data sharing \cite{Abbott2019_LocalStandardsAnonymization, SalehzadehNiksirat2023_ChangesResearchEthics, Waycott2017_EthicalEncountersHCI}. Beyond these professional expectations, data protection laws in many jurisdictions (e.g., the General Data Protection Regulation in most European countries \cite{EuropeanParliament2016_GeneralDataProtection}, the Brazilian Data Protection Law \cite{BrazilianNationalCongress2018_BrazilianDataProtection}, the Personal Data Protection Act in Singapore \cite{ParliamentofSingapore2012_PersonalDataProtection}, and the Personal Information Protection and Electronic Documents Act in Canada \cite{ParliamentofCanada2000_PersonalInformationProtection}) require the protection of research participants’ personal data.

When seeking advice on anonymizing research data before sharing, researchers are likely to encounter the ubiquitous saying "as open as possible, as closed as necessary" \cite{EuropeanResearchExecutiveAgency_OpenScience, Landi2020_FAIROpenPossible}. What this means in practice remains obscure. Additional guidance is mostly example-heavy and therefore extremely case-dependent, without offering researchers procedural advice on how to come to decisions about what closedness is necessary and what openness is possible (e.g., \cite{VanRavenzwaaij2025_DeidentificationWhenMaking}). Well-known privacy concepts such as k-anonymity are often recommended for research data, but their application relies on a series of judgment calls at the threat modeling stage, which are especially uncertain given the complexity of research projects \cite{Guo2025_HowResearchersDeIdentify}.

\begin{figure}
    \centering
    \includegraphics[width=1\linewidth]{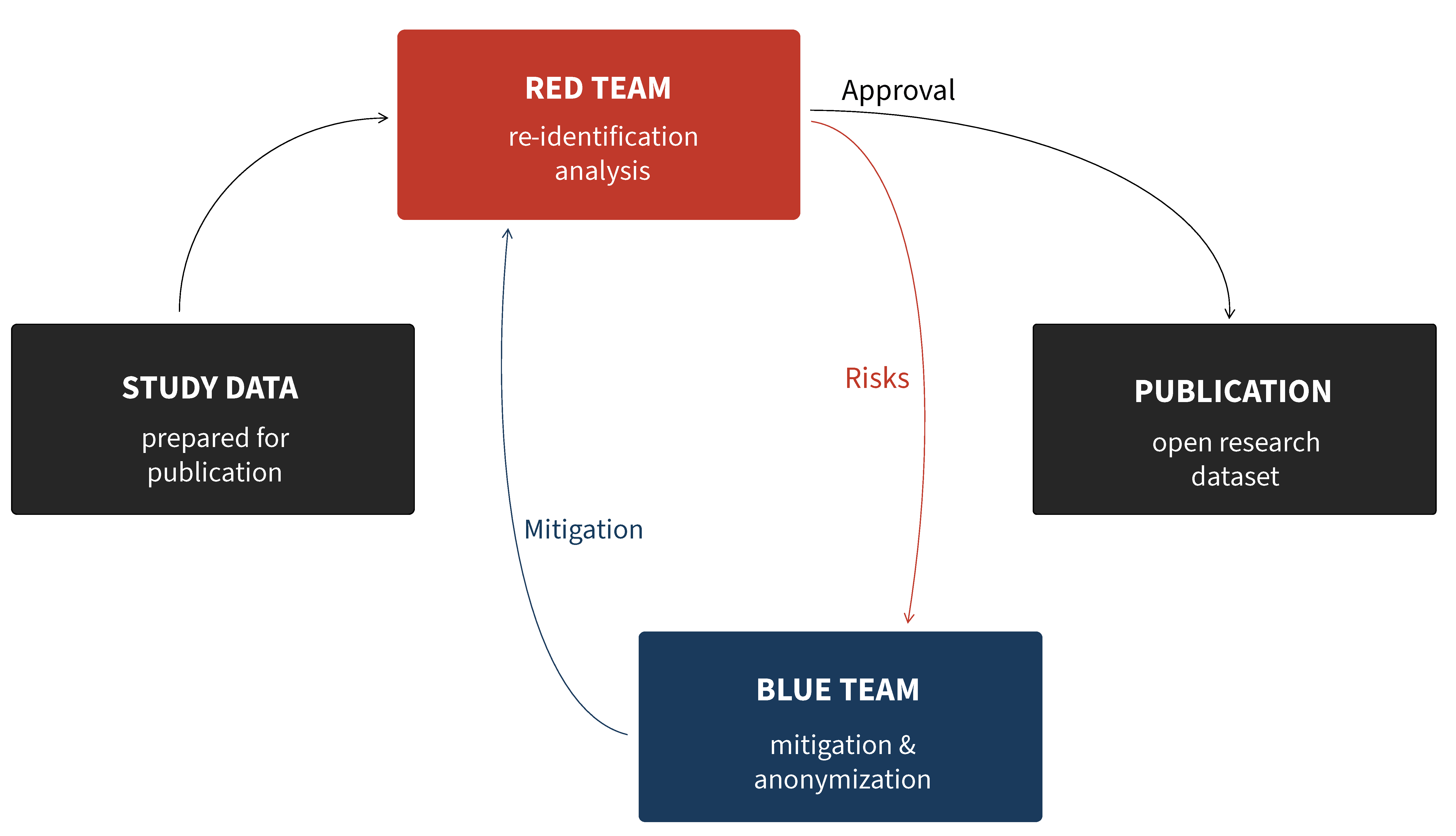}
    \caption{The process of red teaming for testing the robustness of research data anonymization. The process begins with a version of the study data and materials that is only superficially anonymized (e.g., by removing direct identifiers). The red team starts by attacking the anonymization, aiming to re-identify participants. The blue team makes the anonymization more rigorous in reaction to the red team's findings. After one or multiple iterations, when no new risks can be identified, the resulting data and materials are published.}
    \Description{A flowchart of the process of red teaming in research anonymization. The process starts with the study data and materials (minimally anonymized), given to the red team for analysis. They provide the risks identified to the blue team for anonymization. After mitigating these risks, the process may restart. Upon approval, the data are published.}
    \label{fig:Flowchart}
\end{figure}

We therefore propose a procedure to navigate the tradeoff between protecting and opening data: We apply the concept of red teaming vs. blue teaming to determine the actual risks associated with data sharing by attacking our own anonymization (see Figure \ref{fig:Flowchart}). Red teaming has its roots in security practice, initially used for decision-making in military scenarios \cite{Abbass2011_ComputationalRedTeaming}. Nowadays, it is a common method when improving the cybersecurity of organizations \cite{Rehberger2020_CybersecurityAttacksRed, Mansfield-Devine2018_BestFormDefence}. There, the red team is tasked with attacking a specific system, such as an organization's network or physical facilities, behaving in a manner similar to a real adversary \cite{Abbass2011_ComputationalRedTeaming}. The blue team plays their opposition, defending the system and reacting to the red team's attacks \cite{Mansfield-Devine2018_BestFormDefence}. This process is typically iterative, meaning that the red and blue teams respond to each other's actions in multiple rounds or continuously over a specified time period \cite{Mansfield-Devine2018_BestFormDefence}.  

\subsection{Contribution}
Applying this method to the anonymization of research data, we hope to contribute to the privacy efforts of researchers in human-computer interaction (HCI) and elsewhere by making common privacy advice more usable and actionable. We provide materials for researchers to apply the method themselves and demonstrate red teaming for data anonymization with one example from our own research. We aim to spark a discussion on sharing practices of research data and potential methods to balance data utility and data protection in HCI research.

\section{The Method: Red Teaming for Research Data Anonymization}
\subsection{Overview}
We apply red teaming to the anonymization of research data. Initially, the process begins with a research dataset that is only minimally anonymized (e.g., through the suppression of direct identifiers such as names or email addresses), thereby avoiding both trivial re-identification and unnecessary loss of utility. The red team plays the role of the attacker: They are instructed to de-anonymize the research participants. They receive any publicly available information on the study (e.g., recruitment strategy) and may use this information, along with the data, to re-identify any person who participated in the study. They report their results to the blue team. Based on discussions with the red team, the blue team then makes improvements to the anonymization by using anonymization techniques. The newly anonymized data, as well as potentially revised study materials, are then subjected to attack by the red team once again. They attempt to find an alternative method to re-identify the study participants. In case of the red team's success, the blue team can again strengthen the anonymization of the data. Based on the impressions of the red and blue teams, they can proceed with another iteration of the process. See Figure \ref{fig:Flowchart} for an illustration of the process.

The supplementary material (available as a separate document) contains the instruction material on how to utilize this method, including general instructions for the process, tips and tactics for the red team, and tips and techniques for the blue team.

\subsection{The Process}
The red team’s process is very creative. As they play the role of the attacker, they are free to follow any path they deem worthy to re-identify participants. A few starting points for their strategy may include: a systematic exploration of all available materials, including datasets, documentation, and metadata; searching for direct identifiers, pseudonymization errors, and indirect identifiers; linking demographic information, timestamps, language use, procedural metadata, recruitment details with external sources; and even social engineering. Throughout the process, the red team documents all steps taken and reports identified vulnerabilities and plausible re-identification pathways to the blue team.

The red team’s examination can be applied at different stages of the research process. Most effectively, it is conducted prior to data publication in order to assess re-identification risks before datasets or study materials become publicly accessible. Alternatively or even additionally, red teaming can also be conducted after publication. In this case, the red team analyses the publicly available dataset and accompanying materials to identify potential re-identifcation risks that may have been overlooked during the initial anonymization. While post-publication red teaming may be limited in its ability to modify already released data, it can still inform corrective actions such as updating documentation, restricting access to specific variables, or refining anonymization practices in future data releases. An additional consideration when applying red teaming after publication is whether participants should be informed about identified weaknesses in anonymization or newly discovered re-identification risks. 

Both approaches can contribute to strengthening the robustness of research data anonymization. Nevertheless, it should be emphasized that conducting red teaming prior to publication is generally preferable from both a legal and ethical perspective.

The blue team is responsible for mitigating the identified risks, usually following a more structured approach than the red team. Ideally, the blue team members have in-depth knowledge of the dataset, study materials, and planned analyses. Based on the red team’s findings, the blue team improves the anonymization using appropriate techniques, while explicitly considering the trade-off between risks to privacy and data utility. Typical measures include non-perturbative techniques such as recoding, top- and bottom-coding, or suppression; perturbative techniques such as noise addition or data swapping; and de-associative approaches that break links between indirect identifiers and sensitive attributes \cite{Carvalho2023_SurveyPrivacyPreservingTechniques}. In cases where strong privacy protection is necessary, synthetic data generation may be a viable option \cite{Carvalho2023_SurveyPrivacyPreservingTechniques}.

This process is repeated until no further meaningful re-identification risks can be identified or available resources are exhausted. 

\subsection{General Parameters}
One person per team is sufficient to apply the method, provided they are willing to openly acknowledge any weaknesses in the anonymization. The approach is applicable to a wide range of study types and data modalities, as long as the data contain enough information to allow meaningful anonymization decisions.  Throughout the process, careful documentation of attacks, defenses, and resulting changes is essential to ensure transparency and reproducibility of the anonymization strategy.
A key characteristic of the proposed approach is flexibility. Rather than prescribing a rigid protocol, we propose a general framework that allows researchers to adapt the process to the specific research context, community norms, and available resources. As a result, implementations of the method may vary considerably in terms of time invested, personnel involved, and the level of risk that is addressed. 

Anonymization decisions inevitably involve trade-offs between protecting participant privacy and preserving the analytical value of the data. Similarly, the application of red teaming involves balancing the resources invested in the process with the range of potential attack vectors that can realistically be explored. In typical research settings, red teaming is unlikely to identify every possible re-identification strategy or simulate highly sophisticated attacks. However, the approach can substantially reduce the risk that participants can be re-identified with reasonable effort by a motivated adversary. 

To nevertheless provide a brief indication of an effective team composition, certain characteristics of red and blue team members may be beneficial. Members of the red team approach the dataset from the perspective of a potential attacker; therefore, it may be advantageous if they were not directly involved in the original research or anonymization process, as this may reduce confirmation bias and increase the likelihood of identifying overlooked vulnerabilities. At the same time, prior familiarity with the dataset and study materials may help identify concrete weaknesses more quickly. In general, red team members benefit from creativity as well as technical and data-analytic skills that enable them to combine multiple information sources when exploring potential re-identification pathways. 

In contrast, the blue team typically benefits from being composed of members of the core research team who possess detailed knowledge of the dataset, the study design, and the intended analyses. This is especially important when balancing privacy risks against the analytical value of the data during anonymization decisions. Technical expertise in data handling and anonymization techniques may further support the effective mitigation of identified risks. If such expertise is not available within the research team, researchers may consider involving colleagues with relevant data protection or data science expertise or consult institutional experts (e.g., data stewards, research integrity officers) if available. 

\subsection{Legal and Ethical Parameters}
The legal framework applicable to our study is the General Data Protection Regulation (GDPR). The anonymization procedures applied in our study were designed to comply with its principles. 

Whether red teaming is ethically appropriate may be debated, as it deliberately exposes potential vulnerabilities and therefore involves a controlled possibility of re-identification by members of the red team. However, this takes place in a controlled research setting to proactively limit the risk of exposure to a broader audience. Ideally, such a procedure is proactively addressed in the study's consent form. Regarding these challenges, some authors have suggested approaches beyond traditional one-time consent models, in which participants are more directly involved in decisions about anonymization and data governance \cite{Godfrey-Faussett2022_ParticipatoryResearchEthics}.

In our assessment, red teaming does not introduce additional risks, but rather serves as a controlled process to identify vulnerabilities before they can be exploited by external actors. We therefore argue that the approach can contribute both to legal compliance and to the ethical goal of minimizing the risk of participant re-identification. 

\section{Case Report: Applying Red Teaming}
We applied the method ourselves to the data associated with a mixed-method expert study conducted by Luisa Jansen (A1) and others (\href{https://doi.org/10.17605/OSF.IO/8U23P}{link to study and data})\cite{Jansen2023_BridgingGapTraining}. A1 acted as the blue team, while Tim Ulmann (A2) and Robine Jordi (A3) acted as the red team. All red and blue teamers were part of the same research group and access to personal data of the participants was covered by informed consent. Here, we present our process as a case report of the method.

\subsection{The Underlying Study}
The underlying study was a mixed-methods expert study in human-centered privacy. 12 privacy experts were recruited using a freelancing platform. The study evaluated an intervention designed to improve the communication skills of privacy experts, and participants were divided into two study groups. The data included demographics and questions regarding the participants' professional background, ratings of the intervention, and the experts' solutions to communication tasks. Data was collected through both Qualtrics and the intervention platform. To link the data between different sources and surveys, as well as back to the participants, a personal codeword was used. When publishing the study, the aim was to share the data as openly as possible, enabling reuse and verification of the results. 

\subsection{Our Process Red and Blue Teaming}
\subsubsection{Prior Experiences}
All authors of this paper are part of the same research group.
A1 was the primary researcher of the expert study, which serves as the basis for this case report, and possesses conceptual knowledge about privacy and anonymization. A1 did not have any experience with extensive anonymization beyond deleting direct identifiers prior to this process. They usually view anonymization as annoying, but were excited to try this method out. A1 acted as the blue team throughout the whole process. A2 and A3 were not involved in the original case study. Both had basic conceptual knowledge of privacy and anonymization techniques, but no prior practical experience. They both acted as the red team, whereas A3 conducted the first iteration, followed by A2 in the second iteration. They were curious to adopt the attacker's perspective and worked independently without discussing their approaches with each other. 
Malte Elson (A4) was not directly involved in the red teaming process but supervised both the case study and the red teaming process.

\subsubsection{Before the Start of the Red Teaming}
A1, acting as the blue team, minimally anonymized the data by replacing the personal codewords with random numbers that cannot be linked to the participants. 

\subsubsection{Red Teaming: First Iteration}
A1 explained the goal of the red teaming and the overall process to A3, providing A3 with a brief instruction sheet and all available study materials and data. 
A3 started by reverse-engineering the recruitment search strategy on the freelancing platform used in the study and filtering for participants with uncommon demographic characteristics. This led to the identification of profiles for four participants, including their full names. Their participation in the study could be verified through publicly available profile information, in which they listed the study as part of their work experience. Using this information, A3 was able to link these individuals to all other study demographics, including some that were not published in participants' profiles on the freelancing platform, as well as to their performance data. 

This means that information participants intended to keep private (e.g., age or highest level of education) could be inferred from the minimally anonymized data, and that participants' professional skills could be judged based on the performance data in the study. 

\subsubsection{Blue Teaming: First Iteration}
A1 and A3 discussed how to address these issues, including whether to redact the name of the freelancing platform. A1 decided against this, as the platform used for recruiting is central to assessing the validity of the expert study. 

Instead, A1 focused on mitigating the most effective attack vector: identifying participants via their country of residence. To address this, countries were recoded into meaningful categories (i.e., continents and EEA vs. non-EEA within Europe, reflecting the expert study’s focus on EU data protection law). In addition, demographics were de-associated from all other study data to prevent linking individual demographics to performance. Demographic information was further aggregated by reporting summaries for each study group rather than individual-level demographics, reducing the risk of cross-inference between demographic attributes. These latter changes did not meaningfully reduce data utility, as individual relationships between demographics and performance were not of interest; the demographics served primarily to demonstrate comparability between the two groups. Otherwise, perturbative techniques such as swapping values would have been the better option to prevent loss of utility \cite{Carvalho2023_SurveyPrivacyPreservingTechniques}.

\subsubsection{Red Teaming: Second Iteration}
A2 took over the red team from A3 due to personnel changes and to introduce a fresh perspective. A1 explained the goals of the exercise and provided the newly anonymized materials, along with hints about attack paths that had previously been successful for A3, but without disclosing A3’s concrete results. 

A2 again used the freelancing platform to reverse-engineer the recruitment strategy. They found overlooked information in the accompanying materials where participants’ countries of residence were still visible, though without allowing linkage to the other study data. Over approximately eight hours, A2 was able to find the identities of two participants using this information. They could verify their participation in the study by finding references to the study in their public profiles. 

A2 also attempted to use Qualtrics participation timestamps to estimate participants’ time zones in order to link identified individuals to performance data, and tried to match linguistic styles in the study data to public freelancing profiles. Overall, A2 could form tentative inferences based on participation times and linguistic style, but was unable to link any known participant to study data with certainty. This outcome suggests that no definitive re-identification was possible; at most, uncertain guesses could be made about links between individuals and performance data.

\subsubsection{Blue Teaming: Second Iteration}
A1 and A3 discussed the next steps for anonymization. As a result, A1 suppressed dates and times of participation and removed information about participants’ countries of residence from the accompanying materials. After further discussion, they again decided not to redact the name of the freelancing platform.

\subsection{Conclusion}
Working with this method helped us identify the most viable attack vectors and develop countermeasures that rendered them ineffective without sacrificing utility, at least none in terms of verifiability of the expert study's results or the ability to judge the study’s validity. At the same time, the method did not yield definitive answers to every question (e.g., whether the name of the freelancing platform should be anonymized). 

The red team invested a substantial amount of time in understanding the study and navigating the file structure. On the blue-team side, the most time-consuming effort was managing data files and integrating additional anonymization steps into the analysis pipeline, especially since they aimed to publish the anonymization scripts for transparency. Despite this overhead, the process was enjoyable for all of us, turning what is usually an annoying secondary task into an engaging and collaborative one. 

\section{Discussion and Outlook}
Our approach has several limitations. First, red teaming does not eliminate all risks. It does not systematically uncover all possible attack vectors, and its effectiveness depends on the creativity, expertise, and persistence of the team members. 

The process can also be resource-intensive, requiring additional time and effort from the research team. 

Ethical aspects also require further discussion. In particular, one may question whether the potential risk of re-identification by a broader public justifies controlled attempts by a red team to identify participants’ identities. In this context, approaches that go beyond traditional one-time consent models and involve participants more directly in anonymization decisions deserve further consideration. Ideally, participants should at least be informed about and give consent to potential red-teaming procedures as part of the study's consent process. 

Despite these limitations, the approach offers several advantages. While the available resources influence the depth of testing, the red-teaming approach provides a flexible framework that can be adapted to different research contexts and resource levels. Even when not all attack vectors can be explored, the method allows researchers to identify at least the most obvious vulnerabilities and thereby reduce the likelihood of successful re-identification with reasonable effort. In this sense, the approach offers a practical way to operationalize the principle of making data “as open as possible, as closed as necessary.” From both a legal and an ethical perspective, we therefore consider red teaming a useful tool to support responsible data publication. Finally, although anonymization is often perceived as a rather technical or “dry” topic, we experienced the red-teaming process as engaging and enjoyable.

Looking ahead, we plan further testing to better understand for which types of studies this approach is most suitable and whether researchers are willing and able to adopt it in practice. We invite discussions to explore these questions, gather feedback, and refine the method collaboratively.

\bibliographystyle{ACM-Reference-Format}
\bibliography{Data_Protection_of_Research_Data}


\end{document}